%% file: MK-Filter-Arxiv.tex
\documentclass[11pt]{article}
\usepackage{ArxivStyle}

\usepackage[subscriptcorrection]{newtxmath}

\def\idop{{\vmathbb 1}} 

\def\pr{{\mathbb{P}}} 
\newcommand{\E}{E} 

\def\min{{\mathrm{min}}}
\def\max{{\mathrm{max}}}
\usepackage{caption,subcaption}
\usepackage{multirow}
\usepackage{cleveref} 
\usepackage{xcolor} 
\definecolor{darkblue}{rgb}{0,.439,.753}

\usepackage[]{authblk}

\title{Large-scale metric objects filtering for  binary classification with application to abnormal brain connectivity detection}

\author[1]{Shuaida He}
\author[1]{Jiaqi Li}
\author[1]{Xin Chen\footnote{Corresponding author: Xin Chen, chenx8@sustech.edu.cn}}
\affil[1]{Department of Statistics and Data Science, Southern University of Science and Technology, Shenzhen, China}

\begin{document}

\maketitle

\begin{abstract}
	The classification of random objects within metric spaces without a vector structure has attracted increasing attention. 
	However,  the complexity inherent in such non-Euclidean data often restricts existing models to handling only a limited number of features, leaving a gap in real-world applications. 
	To address this, we propose a data-adaptive filtering procedure to identify informative features from a large-scale of random objects,  leveraging a novel Kolmogorov–Smirnov-type statistic defined on the metric space.
	Our method, applicable to data in general metric spaces with binary labels, exhibits remarkable flexibility. 
	It enjoys a model-free property, as its implementation does not rely on any specified classifier.
	Theoretically, it effectively controls the false discovery rate while guaranteeing the sure screening property. 
	Empirically, equiped with a Wasserstein metric, it demonstrates superior sample performance compared to Euclidean competitors.
	When applied to analyze a dataset on autism, our method identifies significant brain regions associated with the condition. 
	Moreover, it reveals distinct interaction patterns among these regions between individuals with and without autism, achieved by filtering hundreds of thousands of covariance matrices representing various brain connectivities.
\end{abstract}

\section{Introduction}
We consider metric binary classification, where features are represented as random objects in general metric spaces. 
Specifically, let $Y \in \{-1,1\}$ be a class label and $X = (X_1,\dots,X_p)$ be an ordered $p$-tuple of features. 
Each $X_j \ (1\leq j \leq p)$ is assumed to be a random object from the metric space $(\mathcal{M}_j,d_j)$, where $\mathcal{M}_j$ is a set, and $d_j$ is the corresponding metric (or distance). 
When $(\mathcal{M}_j,d_j)$ lacks an additional vector structure, it is often referred to as a non-Euclidean space. 
Typically examples include the space of symmetric positive definite (SPD) matrices \citep{lin2019riemannian}, directional objects \citep{mardia2000directional},  phylogenetic trees \citep{billera2001geometry}, subspaces like Grassmannians \citep{lim2021grassmannian}, and probability functions equipped with Wasserstein metric \citep{chen2023wasserstein}. 
Analyzing such non-Euclidean data poses inherent challenges due to the absence of extensive algebraic structures.

Binary classification, given a joint probability measure $\mathbb{P}$ defined over $\mathcal{M} \times \{-1,+1\}$, seeks a function $h: \mathcal{M} \rightarrow \{-1,+1\}$ that minimizes the risk of $\mathbb{P}\{h(X) \neq Y\}$. 
Here, $\mathcal{M}=\prod_{j=1}^{p}\mathcal{M}_j$ denotes the Cartesian product space of $\mathcal{M}_j$, and the function $h$ is usually constrained to a pre-specified class $\mathcal{H}$. 
Commonly employed classes $\mathcal{H}$ include the Lipschitz large-margin classifiers \citep{von2004distance, gottlieb2014efficient}, nearest neighbor classifiers \citep{cover1967nearest, chaudhuri2014rates,kontorovich2018active}, and metric logistic regression models \citep{lin2023logistic}, to name a few. Despite their utility, these metric classifiers often suffer from limited computational efficiency, restricting their applicability to datasets with few features.

In many scenarios, a relatively small sample size $n$ may coexist with a large or even huge value of $p$ (e.g., $p=\exp\{O(n^\alpha)\}$ for some $\alpha>0$), which is kown as the curse of dimensionality in the Euclidean setting.
Nonetheless, it is commonly believed that the underlying classification rule depends only on a few important features $\{X_j, j \in \mathcal{D}\}$, where $\mathcal{D}$ represents a subset of $\{1, \dots, p\}$ with a cardinality $d$ significantly smaller than $n$ (eg., $d=o(n)$), and is termed as the informative set.
This assumption not only facilitates the effective handling of a large-scale of metric features but also provides a deeper understanding of the original classification problem. 
Our primary focus is on identifying such $\mathcal{D}$.

For Euclidean features, an effective solution involves leveraging the celebrated feature screening technique \citep{fan2008sure,fan2018sure}, extensively studied in the literature with diverse variants \citep{zhu2011model,mai2013kolmogorov,cui2015model}. 
Essentially, feature screening aids in constructing a subset of features covering $\mathcal{D}$ with high probability.  
A classic application is locating key genes contributing significantly to tumor classification using high-dimensional micro-array data. 
However, in scenarios where features are random objects in general metric spaces, the identification of $\mathcal{D}$ becomes more challenging and remains relatively unexplored.
To the best of our knowledge, the most promising approach so far is the ball correlation screening procedure \citep{pan2018generic}, designed for objects in separable Banach spaces. Although it can be extended to metric spaces according to the Banach-Mazur theorem \citep{kleiber1969generalized}, this extention is less straightforward and may loss some power in a binary classification setting.

In this paper, we estimate the informative set $\mathcal{D}$ by developing an effective filtering procedure that operates on a large set of random objects in metric spaces with binary labels. 
This filtering procedure is designed to avoid relying on strong model assumptions, making it applicable to arbitrary classifier classes $\mathcal{H}$. 
Additionally, it exhibits flexibility to handle various metric spaces. 
To achieve this, we leverage a Kolmogorov–Smirnov-type statistic defined on the metric space. 
Our filtering procedure, based on this novel statistic, extends the approach proposed in \citet{mai2013kolmogorov} to the non-Euclidean setting.
To balance classification performance and computational efficiency, the proposed method filters out as many irrelevant objects as possible while retaining informative features. 
We demonstrate this by showing that it enjoys the sure screening property while effectively controlling the false discovery rate, when combined with a data-adaptive threshold selection procedure. 

As another contribution of this work, we apply the proposed method to detect abnormal functional connectivities in autism brains. 
We first construct a large pool of symmetric positive definite matrices as metric features to characterize various brain functional connectivities and then apply our filtering method. 
The identified informative features are highly interpretable, revealing significant differences in brain connectivity patterns between individuals with autism and controls. 
Some of these findings are less reported in existing autism studies and may warrant future investigation. 
Moreover, in the subsequent classification task, the classifier based on these informative features demonstrates significantly enhanced prediction performance under various criteria.

Besides binary classification, recent interest in non-Euclidean data analysis has also spurred the development of many other methods. These encompass Fr{\'e}chet mean estimation in metric spaces \citep{mccormack2021equivariant}, Fr{\'e}chet linear regression \citep{petersen2019frechet}, Fr{\'e}chet single index models \citep{bhattacharjee2023single}, sufficient dimension reduction for random objects \citep{ying2022frechet, zhang2023dimension},  functional models for time-varying random objects \citep{dubey2020functional}, and more.  Again, despite their wide range of applications, these methods are not specifically designed to handle a large-scale of non-Euclidean features.

\section{Preliminaries}
\subsection{Examples of metric spaces}\label{sec:ws}
We briefly introduce some metric spaces that will be encountered later in the paper. 
\begin{example}[The Wasserstein space of univariate distributions] 
	Let $(\mathcal{X},\mathcal{B}(\mathcal{X}))$ be a measurable space, where $\mathcal{X}$ is either $\mathbb{R}$ or a closed interval of $\mathbb{R}$, and $\mathcal{B}(\mathcal{X})$ is the associated Borel $\sigma$-field. 
	The Wasserstein space $\mathcal{W}(\mathcal{X})$ is the set of probability distributions $F$ on $(\mathcal{X},\mathcal{B}(\mathcal{X}))$ such that $\int_{\mathcal{X}}x^2 d F(x) < \infty$, equipped with quadratic Wasserstein distance $d_{w}(F_1,F_2)=(\int_{0}^{1} [F_1^{-1}(s) - F_2^{-1}(s)]^2 ds)^{1/2}$, 
	where $F_1, F_2 \in \mathcal{W}$ and $F_1^{-1}$ and $F_2^{-1}$ are the quantile functions corresponding to $F_1$ and $F_2$, respectively. 
	Then $(\mathcal{W}(\mathcal{X}),d_w)$ is a metric space with a Riemannian structure \citep{chen2023wasserstein}. 
\end{example}

\begin{example}[The space of  SPD matrices with a fixed dimension]
	Let $\mathcal{S}_m^{+}$ denote the set of $m \times m$ symmetric positive-definite (SPD) matrices. 
	Generally, $\mathcal{S}_m^{+}$ forms a smooth submanifold embedded in $\mathbb{R}^{m(m+2)/2}$. 
	When equipped with the inherited Euclidean metric, it becomes a Riemannian manifold.
	However, the determinant of Euclidean average of SPD matrices may larger than any of the original determinants, leading to the well-known phenomenon of swelling effect \citep{arsigny2007geometric}. This limitation, along with others, motivates the development of various alternative metrics for $\mathcal{S}_m^{+}$, such as the Log-Euclidean metric, the Cholesky distance \citep{10.1214/09-AOAS249}, and the Log-Cholesky \citep{lin2019riemannian}, among others. See illustrative examples in section \ref{sec:simu-spd}.
\end{example}

\subsection{Metric distribution function}
Suppose $X$ is a random object taking values in a  general metric space $(\mathcal{M},d)$ and $\mu$ is the associated Borel probability measure. 
Throughout this article, we assume $(\mathcal{M},d)$ is a complete and separable (i.e., Polish) space without the linear structure. 
To perform nonparametric statistical inference on samples from this non-Euclidean space $\mathcal{M}$, \citet{wang2023nonparametric} introduced a fundamental tool termed metric distribution function (MDF).
They proved that MDF, as a quasi-distribution function, uniquely determines $\mu$ when $(\mathcal{M},d)$ is a Polish space and $d$ satisfies some mild conditions;
see Theorem 1 in \citet{wang2023nonparametric} for details. 
This key result suggests that MDF can serve a role in metric space-based statistical inference akin to the distribution function in the Euclidean setting.
We provide a brief review of MDF before moving on.

Roughly speaking, the metric distribution function is constructed upon all open balls in $(\mathcal{M},d)$, which is a base of the metric topology. 
Denote $B(u,r)=\{v:d(u,v)< r\}$ as the open ball in $(\mathcal{M},d)$ where $u$ is the center and $r \geq 0$ is the radius. 
Let $\bar{B}(u,r)=\{v:d(u,v)\leq r\}$ be the corresponding closed ball. 
For $\forall u, v \in \mathcal{M}$, the metric distribution function of $\mu$ on $\mathcal{M}$ is defined as 
\[F^{M}(u,v)=\mu\left[\bar{B}\{u, d(u,v)\}\right] = \E\{\delta(u,v,X)\},\]
where $\delta(u,v,x) =  \idop\left[x \in \bar{B}\{u, d(u,v)\}\right]$, $\idop(\cdot)$ is the indicator function, and the superscript $M$ indicates the metric space.
Based on i.i.d. samples $\{X_1,\dots, X_n\}$ from $\mu$ on $\mathcal{M}$, a natural estimate of MDF is the empirical metric distribution function (EMDF) 
\[F_{n}^{M}(u,v) =\frac{1}{n}\sum_{i=1}^{n}\delta(u,v,X_i).\]
The following lemma from \citet{wang2023nonparametric} establishes the uniform convergence result for EMDF, indicating it enjoys a Glivenko-Cantelli-type property. 
\begin{lemma}[Theorem 3 in \citet{wang2023nonparametric}]
	\label{th:lem1}
	Let $\mathcal{F}=\{\delta(u,v,\cdot): u, v \in \mathcal{M}\}$ be the collection of indicator functions of closed balls on $\mathcal{M}$. 
	Let $\{X_1,\dots,X_n\}$ be i.i.d. samples from $\mu$. 
	Define $X_1^{n}=(X_1,\dots,X_n)$ and $\mathcal{F}(X_1^{n})=\{(f(X_1),\dots,f(X_n))\mid f \in \mathcal{F}\}$.
	If $\mu$ satisfies that $\frac{1}{n} \E_{X}[\log(\mathrm{card}(\mathcal{F}(X_1^n)))] \rightarrow 0$, where $\mathrm{card}(\cdot)$ is the cardinality of a set, then
	\[\lim_{n \rightarrow \infty} \sup_{u \in \mathcal{M}, v \in \mathcal{M}}\left|F_{\mu, n}^{M}(u, v)-F^{M}(u, v)\right|=0,\ a.s.\]
\end{lemma}


\section{The Metric Kolmogorov Filter}
\subsection{Method}
Let $X_j \mid Y=+1 \sim \mu_{j}^+$ and $X_j \mid Y=-1 \sim \mu_{j}^-$, where $\mu_{j}^+$ and $\mu_{j}^-$ represent two unknown Borel probability measures on the metric space $(\mathcal{M}_j, d_j)$. 
Define $F^{M}_{+j}(u,v)$ as the metric distribution function of $d_{j}(u,v)$ for $u, v \sim \mu_{j}^{+}$, and similarly, $F^{M}_{-j}(u,v)$ for $u, v \sim \mu_{j}^-$.
Apparently, a significant distinction between $\mu_{j}^+$ and $\mu_{j}^-$ implies $X_j$ contributes to predicting label $Y$. 
Due to the one-to-one correspondence between a probability measure and its metric distribution function, we are motivated to consider the following statistical divergence 
\[\operatorname{MKS}(\mu_{j}^{+} \| \mu_{j}^{-})=\int_{\mathcal{M}_j} \sup_{v \in \mathcal{M}_j} \mid F^{M}_{+j}(u,v) - F^{M}_{-j}(u,v)\mid d \mu^{+}(u)\]
to distinguish $\mu_{j}^+$ from $\mu_{j}^-$.
It works by first measuring the Kolmogorov-Smirnov divergences between $F^{M}_{+j}(u,v)$ and $F^{M}_{-j}(u,v)$ for each fixed ball center $u \sim \mu_j^+$, then integrates them by expectation.
This statistic is termed the Metric Kolmogorov-Smirnov (MKS) divergence,  which has been first visited in a manuscript of  \citet{wang2021nonparametric} for homogeneity test. 
Here we symmetrize MKS to measure the importance of $X_j$ to $Y$ by considering the following nonnegative statistic 
\begin{equation}\label{fm:wj}
	\omega_j =  \operatorname{MKS}(\mu_{j}^{+} \| \mu_{j}^{-}) + \operatorname{MKS}(\mu_{j}^{-} \| \mu_{j}^{+}).
\end{equation}
Intuitively, $\omega_j $ quantifies the marginal association of $X_j$ and $Y$, that is, the larger $\omega_j$ is,  the more important $X_j$ is.
Conversely, when $\omega_j$  tends to zero, it indicates an independent relationship between $X_j$ and $Y$, as shown by Proposition~\ref{th:prop-ind}. 
\begin{proposition}\label{th:prop-ind} 
	$X_j$ and $Y$ are statistically independent if and only if $\omega_j=0$.
\end{proposition}

At the sample level, we estimate $\omega_j$ based on the empirical metric distribution. 
For ease of presentation, let $\mathcal{X}^+$ be a dataset containing $n_1$ i.i.d. samples with positive labels.
Thus for $i \in \{1,\dots,n_1\}$ and $j \in \{1,\dots,p\}$, the $i$th random tuple is denoted as $X_i^{+} \in \mathcal{X}^+$, with the $j$th object $X_{i,j}^+ \sim \mu_j^{+}$.
Similarly, denote $\mathcal{X}^-$ as a dataset containing $n_2$ samples with negative labels.
The complete dataset is then $\mathcal{X}=\mathcal{X}^{+} \cup \mathcal{X}^{-}$, and the total sample size $n=n^{+}+n^{-}$.
We utilize 
\[\widehat{\operatorname{MKS}}(\mu_j^{+} \| \mu_j^{-}) = \frac{1}{n^+} \sum_{i=1}^{n^+} \max_{v \in \mathcal{X}}\left|F_{+j, n^+}^{M}(X_{i,j}^{+}, v) - F_{-j, n^-}^{M}(X_{i,j}^{+}, v)\right|\]
and 
\[\widehat{\operatorname{MKS}}(\mu_{j}^{-} \| \mu_{j}^{+})=\frac{1}{n^-} \sum_{i=1}^{n^-} \max_{v \in \mathcal{X}}\left|F_{-, n^-}^{M}(X_{i,j}^{-}, v) - F_{+,  n^+}^{M}(X_{i,j}^{-}, v)\right|\]
to estimate $\omega_j$, that is, 
\begin{equation}\label{fm:wnj}
	\widehat{\omega}_j = \widehat{\operatorname{MKS}}(\mu_j^{+} \| \mu_j^{-})+\widehat{\operatorname{MKS}}(\mu_j^{-} \| \mu_j^{+}).
\end{equation}

Now we define $\mathcal{S}=\{j:\omega_j > 0, j=1,\dots,p\}$ as the selected set that contains all important features, with cardinality $s$. 
In binary classification, we further suppose $\mathcal{S}$ equals to the informative set $\mathcal{D}$, or more conservatively, $\mathcal{D} \subseteq \mathcal{S}$. 
This assumption is essential for our  theoretical study and aligns with similar assumptions in Euclidean screening literature; see Theorem 1 in \citet{mai2013kolmogorov} for reference.
To estimate $\mathcal{S}$, we rank all random objects $X_j$ based on $\widehat{\omega}_j$ and select those among the $\hat{s}$th largest, where a theoretical choice of $\hat{s}$ is the integer $\ceil{n/\log(n)}$.
Equivalently, we can also specify a threshold $T$ to filter out uninfomative features, resulting 
\begin{equation}\label{fm:info-set}
	\widehat{\mathcal{S}}(T)=\{j: \widehat{\omega}_j\geq T,\ \mathrm{for}\ 1\leq j \leq p  \}.
\end{equation}
In this case, we slightly abuse notation by using $\hat{s}(T)$ to represent the size of $\widehat{\mathcal{S}}$ when employing a threshold $T$.  We term the above procedure the Metric Kolmogorov filter (MK-Filter), considering $\widehat{\omega}_j$ is constructed based on a symmetrized metric Kolmogorov-Smirnov divergence.

\subsection{Sure screening property} 
The MK-Filter is designed to identify the underlying informative set with high probability. 
We demonstrate this by showing that the MF-Filter enjoys the sure screening property, a key characteristic commonly shared by various Euclidean screening techniques. 
To establish the theoretical groundwork for subsequent discussions, we first introduce the following lemma, which provides a uniform concentration inequality for the EMDF.
\begin{lemma}[A uniform concentration inequality]
	\label{th:lem-ineq}
	Let $\mu$ be a probability measure defined on the metric space $(\mathcal{M},d)$ and $\{X_1,\dots,X_n\}$ be i.i.d. observations sampled from $\mu$. 
	For each $\epsilon>0$, there exists a universal constant $N(\epsilon) \in \mathbb{N}$ such that for all $n \geq N(\epsilon)$, we have 
	\[\pr \left\{ \frac{1}{n}\sum_{i=1}^{n} \sup_{v\in\mathcal{M}} \left|F_{\mu,n}^{M}(X_i, v) - F_{\mu}^{M}(X_i, v)\right|>\epsilon \right\}  \leq 2\exp(-\frac{n\epsilon^2}{32}).\]
\end{lemma}

Lemma~\ref{th:lem-ineq} extends the classical Dvoretzky–Kiefer–Wolfowitz inequality \citep{dvoretzky1956asymptotic} to metric spaces, showing that the EMDF exhibits uniform concentration over the sample set at an exponential rate. 
This finding parallels Corollary 3 in \citet{wang2023nonparametric}, but instead of taking the supremum, we compute the sample average across various ball centers, thus leading to a slightly improved bound.
Using this result, Lemma~\ref{th:lem-w-converge} establishes the consistency of $\widehat{\omega}_j$, followed by demonstrating the sure screening property in Theorem~\ref{th:sure}.

\begin{lemma}\label{th:lem-w-converge}
	Under the conditions of \Cref{th:lem-ineq}, for $\widehat{\omega}_j$ defined by \eqref{fm:wnj}, 
	we have $\pr (|\widehat{\omega}_j - \omega_j| \geq c n^{-\kappa}) \leq O (\exp\{-Cn^{1-2\kappa}\})$
	with $0<\kappa<1/2$.
\end{lemma}

\begin{theorem}[Sure screening property]\label{th:sure}
	Define $\xi(\epsilon)= 4\exp(- \frac{n\pi_{+}\epsilon^{2}}{256}) + 2\exp(-\frac{n\pi_{-}\epsilon^{2}}{256})+ \exp(-\frac{c_{1}n\pi_{+}^{2}}{4}) + \exp(-\frac{c_{2}n\pi_{-}^{2}}{4})$, where $\pi_{y}=\pr(Y=y)$ for $y=+1, -1$.
	Assume there exists a selective set $S$ such that the informative set $D \subseteq S$. 
	Let $\delta_{S}= \min_{j\in S}\ \omega_j - \max_{j\in S^{c}}\ \omega_j >0$. 
	Then we have $\pr\{D \subseteq \widehat{S}(d_n)\} \geq 1 - p\xi(\delta_{S}/2)$. 
	Thus, if $\delta_{S} \gg \{\log(p)/n\}^{1/2}$, the sure screening property holds with probability going to 1.
\end{theorem}

\subsection{A data-adaptive threshold for FDR controlling}
The MK-Filter yields a selected set $\widehat{\mathcal{S}}(T)$ with a given threshold $T$, where $T$ controls the set size and, consequently, influences the filtering procedure's quality. 
Generally, a conservative strategy tends to choose a small value of $T$ to include all informative features  with high probability. While ensuring the sure screening property, this strategy also elevates the risk of retaining more uninformative predictors. 
As a more reliable approach, we propose controlling the false discovery rate (FDR, \citet{benjamini1995controlling}) for MK-Filter through a data-adaptive threshold selection procedure. 
Our approach is inspired by \citet{guo2023threshold} but is adapted for handling Non-Euclidean features.

Mathematically, $\widehat{\mathcal{S}}(T)$ makes a false discovery if there exists a $j \in \widehat{\mathcal{S}}(T) \bigcap \mathcal{S}^{c}$, where $\mathcal{S}^{c}$ denotes the underlying true uninformative set. 
The false discovery rate is then defined as 
\[\E[\mathrm{FDP}\{\widehat{\mathcal{S}}(T)\}],\]
where 
\[\mathrm{FDP}\{\widehat{\mathcal{S}}(T)\}=\frac{\#\{j:  j \in \widehat{\mathcal{S}}(T) \bigcap \mathcal{S}^{c}\}} {\#\{j: j \in \widehat{\mathcal{S}}(T)\}},\]
stands for the false discovery propotion (FDP). 
With a pre-specified level $\alpha$, we say using threshold $T$ controls the FDR asymptotically if 
$\lim\sup_{n \rightarrow \infty} \E[\mathrm{FDP}\{\widehat{\mathcal{S}}(T)\}] \leq \alpha.$

We take a data splitting strategy to determine the threshold $T$ for FDR control. 
Firstly, the whole dataset $\mathcal{X}$ is randomly divided into disjoint parts $\mathcal{X}_1$ and  $\mathcal{X}_2$ of different sample sizes $n_1 =n(K-1)/K$ and $n_2 = n/K$, where $K \geq 3$ and $n/K$ are assumed to be integers for simplicity. 
Let $\widehat{\omega}_{j1}$ and $\widehat{\omega}_{j2}$ be the symmetrized Metric Kolmogorov-Smirnov statistics calculated on $\mathcal{X}_1$ and  $\mathcal{X}_2$ respectively. 
Define 
\begin{equation}
	\label{fm:W}
	W_j = \mathrm{sign}(n_1^{\gamma}\widehat{\omega}_{1j} - n_2^{\gamma}\widehat{\omega}_{2j})\max(n_1^{\gamma}\widehat{\omega}_{1j}, n_2^{\gamma}\widehat{\omega}_{2j}),
\end{equation}
then a data-adaptive threshold is obtained via 
\begin{equation}
	\label{fm:threshold}
	T=\inf\left\{t>0: \frac{1+\#\{j: W_j \leq -t\}}{\max(\#\{j: W_j \geq t\},1)}\leq \alpha \right\}.
\end{equation}
Combined with this data-adaptive threshold selection procedure, the MK-Filter works as follows.

\begin{algorithm}
	\caption{The MK-Filter with a data-adaptive threshold.}
	\label{alg}
	1. Randomly divide the dataset $\mathcal{X}$ into disjoint parts $\mathcal{X}_1$ and $\mathcal{X}_2$;\\
	2. Calculate $\widehat{\omega}_{1j}$ and $\widehat{\omega}_{2j}$ on $\mathcal{X}_1$ and $\mathcal{X}_2$ separately using (\ref{fm:wnj}) for  $j = 1,\dots,p$;\\
	3. Obtain $W_j$ via (\ref{fm:W}) and choose the threshold $T$ by (\ref{fm:threshold});\\
	4. Output $\widehat{\mathcal{S}}(T)=\{j: W_j \geq T\}$.
\end{algorithm}

Prior to delving into the details of our threshold selection procedure, we present a short discussion on the statistic $W_j$ defined in formula (\ref{fm:W}) , which distinguishes between the behaviors of informative and uninformative predictor $X_j$. 
Specifically, for $j \in \mathcal{S}$, by Lemma~\ref{th:lem-w-converge} we have $\widehat{\omega}_j \overset{p}\to \omega_j > 0$, thus $W_j >0$ with probability tending to one, given that $n_1^{\gamma}\widehat{\omega}_{1j} > n_2^{\gamma} \widehat{\omega}_{2j} >0$. 
Conversely, for $j \in \mathcal{S}^c$, $W_j$ is asymptotically symmetric about zero due to the independence of $\mathcal{X}_1$ and $\mathcal{X}_2$. 
Since this marginal symmetry property holds for all uninformative predictors, we can use $\#\{j: W_j \leq -t\}$ to approximate both $\#\{j \in \mathcal{S}^c: W_j \leq -t\}$ and $\#\{j \in \mathcal{S}^c: W_j \geq t\}$.
Additional theoretical insights into $W_j$ and $T$ are available in \citet{guo2023threshold}.

We demonstrate that Algorithm~\ref{alg} enables the MF-Filter to control the false discovery rate asymptotically under mild conditions. 
For clarity, we introduce following notations. 
Let $\delta_{n}=c n^{-\kappa}$ and $\xi(\delta_n)=\exp(-Cn\delta_{n}^2)$, where constants $c>0$, $C>0$, and $0<\kappa<1/2$.
Define $\mathcal{C}_{\beta}=\{j \in \mathcal{S}: \omega_{j} > 2\delta_{n}  a/(a-1) \}$, with $\beta$ representing the cardinality of $\mathcal{C}$ and $a=(K-1)^{1/2}$. 
Conceptually, the set $\mathcal{C}_{\beta}$ collects predictors with identifiable signal strengths.
We also define $G_{+}(t)=(p-s)^{-1}\sum_{j \in \mathcal{S}^c}\pr(W_j \geq t)$ and $G_{-}(t)=(p-s)^{-1}\sum_{j \in \mathcal{S}^c}\pr(W_j \leq -t)$.
The following technical conditions, introduced in \citet{guo2023threshold}, are required to guarantee the result in proposition~\ref{th:prop-fdr}. Intuitively, the fraction in Condition~\ref{condition:1} estimates the false discovery proportion, while the Condition~\ref{condition:1} characterizes the convergence of emprical distribution $\{(p-s)G_{+}(t)\}^{-1}\sum_{j\in \mathcal{S}^c}\idop(W_j\geq t)$ and $\{(p-s)G_{-}(t)\}^{-1}\sum_{j\in \mathcal{S}^c}\idop(W_j\leq -t)$.

\begin{condition}\label{condition:1}
	For  $j\in \mathcal{S}^c$, we have $n^{1/2}\widehat{\omega}_j \overset{d}\to \mathcal{N}_j$ and $\pr(n^{1/2}\widehat{\omega}_j > t)/(1-F_{\mathcal{N}_j})\rightarrow 1$ uniformly in $j \in \mathcal{S}^c$ and $t \in (0,n^{1/2})$, as $n\rightarrow \infty$ and $t\rightarrow \infty$,   where $\mathcal{N}_j$ is some nondegenerate random variable with distribution function $F_{\mathcal{N}_j}$. 
\end{condition} 

\begin{condition}\label{condition:2}
	For $\beta \rightarrow \infty$, we have 
	\[\sup _{0 \leq t \leq G_{+}^{-1}\left(\alpha \beta / p\right)}\left|\left\{(p-s)G_{+}(t)\right\}^{-1} \sum_{j \in \mathcal{S}^c} \idop\left(W_j \geq t\right)-1\right|=o(1)\]
	and 
	\[\sup _{0 \leq t \leq G_{-}^{-1}\left(\alpha \beta / p\right)}\left|\left\{(p-s)G_{-}(t)\right\}^{-1} \sum_{j \in \mathcal{S}^c} \idop\left(W_j \leq -t\right)-1\right|=o(1).\]	
\end{condition}

\begin{proposition}\label{th:prop-fdr}
	Assume $p=\exp\{o(n^{1-2\kappa})\}$ and Condition~\ref{condition:1}-\ref{condition:2} holds.
	If $\beta \rightarrow \infty$, $(p-s)\xi(\delta_n) \rightarrow 0$ and $(1-1/K)^{1/2}\delta_{n} \rightarrow 0$ as $(n,p)\rightarrow\infty$, then  Algorithm \ref{alg}  yields a set $\mathcal{S}(T)$ that ensures 
	$\lim\sup_{(n,p)\rightarrow\infty} \mathrm{FDR}\leq \alpha$ for any $\alpha \in (0,1)$. 
\end{proposition}

\section{Simulations}
\input{doc/simulation.tex}

\section{Application: abnormal connectivity detection in autism brains}\label{sec:real-data}
\input{doc/real-data.tex}

\section{Discussion}
In this paper, we proposed a filtering method designed to identify informative features for binary classification from a large set of non-Euclidean objects in metric spaces. 
We conduct a comprehensive analysis, exploring both its theoretical properties and application performance. 
While our discussion primarily focuses on its effectiveness with SPD matrices, our method's applicability can be extended to handle a broader range of metric features.

It is worth noting that the effectiveness of our approach hinges on the choice of the metric used to differentiate between various random objects. Selecting an appropriate metric is an essential problem of independent interest, closely related to the application background, and should be justified case by case.

\bibliographystyle{plainnat}
\bibliography{paper-ref}

\end{document}

%% file: doc/simulation.tex
\subsection{Distributional predictors}
It is interesting to compare the MK-Filter with existing Euclidean screening methods. 
Consider a feature vector $X=(X_1,\dots,X_p)^\T \in \mathbb{R}^p$ and a label $Y \in \{-1,+1\}$. 
In this scenario, classical methods such as Kolmogorov-Filter  \citep{mai2013kolmogorov}, MV-SIS \citep{cui2015model}, DC-SIS \citep{zhu2011model}, and Ball-SIS \citep{pan2018generic} can be readily applied. 
Meanwhile, applying the MK-Filter necessitates specifying a suitable metric space first. 
A reasonable choice is the Wasserstein space $(\mathcal{W}, d_w)$ of univariate distributions, as described in Section \ref{sec:ws}. 
Here, for each $i \in \{1,\dots,n\}$ and $j \in \{1,\dots,p\}$, we denote the $i$th sample of the $j$th feature with the label $Y_i=y$ as $X_{i,j}^y$. Then, its distribution function $F_{i,j}^y$ can be treated as an object in $(\mathcal{W}_j, d_w)$.

As discussed earlier, the MK-Filter identifies the informative set  by ranking $\widehat{\omega}_j$ for each feature $X_j$, which relies on estimating the metric distribution functions $F_{+j,n^+}^{M}$ and $F_{-j,n^-}^{M}$. 
A crucial step in this process involves computing Wasserstein distances $d_{w}(F_{i,j}^y, F_{i^{\prime},j}^{y^\prime})$ for pairs of random objects $F_{i,j}^y$ and $F_{i^{\prime},j}^{y^\prime}$. 
In practice, however, these $F_{i,j}^y$ and  $F_{i^{\prime},j}^{y^\prime}$ are distribution functions that are often not fully observed, requiring estimation. 
A straightforward yet effective approach is to use the empirical distribution functions. 
To be specific, we use $m_{yj}$ observations  $\{X_{i,j}^{l}\}_{l=1}^{m_{yj}}$ with label $y$ to obtain the estimate $\widehat{F}_{i,j}^{y}$, where  $\widehat{F}_{i,j}^{y}(t)=\sum_{l=1}^{m_{yj}}\idop_{(-\infty,t]}(X_{i,j}^{l})/m_{yj}$ for all $ t \in \mathbb{R}$. 
We then estimate $d_{w}(F_{i,j}^y, F_{i^{\prime},j}^{y^\prime})$ by $d_{w}(\widehat{F}_{i,j}^{y}, \widehat{F}_{i^{\prime},j}^{y^{\prime}})$, see \citet{fournier2015rate} and \citet{lei2020convergence} for the theoretical justification. 
Without loss of generality, we further set $m_{+j}=m_{-j}=m$, simplifying the computation to 
$d_{w}(\widehat{F}_{i,j}^y, \widehat{F}_{i^{\prime},j}^{y^\prime}) = [\sum_{l=1}^{m}\{X_{i,j}^{(l)} - X_{i^{\prime},j}^{(l)}\}^2]^{1/2}$. 
Here, $X_{i,j}^{(l)}$ and $X_{i^\prime,j}^{(l)}$ represent the $l$th order statistics of the sample sets $\{X_{i,j}^{l}\}_{l=1}^{m}$ and $\{X_{i^\prime,j}^{l}\}_{l=1}^{m}$, respectively.

We conside the following model in our simulatoin. 
Suppose $Y$ is generated from the discrete uniform distribution with two categories $\{+1, -1\}$
where $\pr(Y = +1) = 1/2$. 
Given $Y=y$, the $p$-dimensional feature vector $X=(X_{1},\dots,X_{p})^\T$ is generated as follows.
\begin{enumerate} 
	\item  $X_{1}^+ \sim N(0.3,1)$ and $X_{1}^- \sim N(-0.3,1)$;
	\item  $X_{2}^+ \sim \mathrm{Uniform}(-1,1)$ and $X_{2}^- \sim \mathrm{Uniform}(-0.8,1.2)$;		
	\item  $X_{3}^+ \sim N(0,1)$ and $X_{3}^- \sim N(0,1.5)$;
	\item $X_{4}^+ \sim \mathrm{Uniform}(-1,1)$ and $X_{4}^- \sim \mathrm{Uniform}(-1.4,1.4)$;		
	\item  $X_{5}^+ \sim N(0,1)$ and $X_{5}^- \sim t(3)$;	
	\item  $X_{6}^+ \sim t(3)$ and $X_{6}^- \sim t(1)$;

	\item  $X_{7}^+ \sim \mathrm{GEV}(\mu,\sigma_1,\xi)$ and $X_{7}^- \sim \mathrm{GEV}(\mu,\sigma_2,\xi)$, where $\mu=0$, $\sigma_1=0.1$, $\sigma_2=0.2$, and $\xi=0$;	
	\item  $X_{8}^+ \sim \mathrm{GEV}(\mu,\sigma,\xi_1)$ and $X_{8}^- \sim \mathrm{GEV}(\mu,\sigma,\xi_2)$, where $\mu=0$, $\sigma=0.1$, $\xi_1=0.1$, and $\xi_2=0.4$;					
	\item Both $X_{j}^+$ and $X_{j}^- \sim N(0,1)$ for $j \in \{9,\dots,p\}$.
\end{enumerate}
Clearly, the first  $8$ features are informative. 
Setting 1-2, for example, characterize the mean differences between the distribution of $X_j^+$ and $X_j^-$.
Setting 3-4 represent instances with the identical means but varying degrees of variation. 
Setting 5-6 stress heavy-tailed cases, comparing Student's t-distributions with different degrees of freedom.
Lastly, Setting 7-8 explore generalized extreme value distributions,  varying in scales (controlled by $\sigma$) or shapes (controlled by $\xi$).

We compare our MK-Filter with the Kolmogorov-Filter  \citep{mai2013kolmogorov}, MV-SIS \citep{cui2015model}, DC-SIS \citep{zhu2011model}, and Ball-SIS \citep{pan2018generic}. 
To handle a feature dimension of $p=10000$, the MK-Filter employs $m=20$ samples to compute each $\widehat{F}_{i,j}^y$, resulting in a dataset of $n=40$ random objects. 
To ensure fair comparision, all other Euclidean screening methods use $n \times m=800$ samples for their estimation. 
To evaluate the effectiveness of each method, we compute the minimum model size (MMS, \citet{fan2010sure}) to include all informative features. 
Additionally, we compute the proportion $\mathcal{P}_{j}^{s}$ of a single informative feature $X_j$ within a selected set of size $s$,  across 400 repeated experiments. 
Table~\ref{mkf:table1} summarizes the quantiles of MMS, along with $\mathcal{P}_j^{s}\ (j = 1, \dots, 8)$ for a given model size $s = \ceil{n/ \log n}$, where $\ceil{x}$ represents the integer part of $x$. 
\input{doc/table-1.tex}

We can see from Table~\ref{mkf:table1} that all methods demonstrate excellent performance in detecting mean differences. 
However, the Kolmogorov-Filter and MV-SIS show less effectiveness when handling normal distributions with the same mean but slightly different variances, while most methods struggle in heavy-tailed settings. Our method consistently exhibits screening efficiency comparable to or better than others across all scenarios. 
This positive outcome can be reasonably attributed to the effective characterization of subtle distributional differences by the Wasserstein distance. 
Our MK-Filter, designed for use in metric spaces, seamlessly integrates with Wasserstein distance, resulting in satisfactory performance across various distributional differences.

\subsection{Symmetric positive definite matrices}\label{sec:simu-spd}
We consider a scenario where features consist of symmetric positive definite matrices. 
Given the label $Y=y$, assume each random object $X_j$ of the $p$-tuple features $X=(X_{1},\dots,X_{p})$ follows a Wishart distribution,
denoted as $X_j^y \sim \mathcal{W}_{m}(v_m,\Sigma_y) \in \mathbb{R}^{m \times m}$, where $v_m>m-1$ represents the degree of freedom, and $\Sigma_y \succ 0$ is the scale matrix. 
We generate $Y$ from the discrete uniform distribution with $\pr(Y = +1) = \pr(Y = -1)= 1/2$, then draw $X_j$ from the following models.
\begin{enumerate}
	\item  For $j\in \{1, 2\}$, $X_{j}^+ \sim \mathcal{W}_{m}(v_m, I_m)$ and $X_{j}^- \sim \mathcal{W}_{m}(v_m, 0.6*I_m)$;
	\item  For $j\in \{3, 4\}$, $X_{j}^+ \sim \mathcal{W}_{m}(v_m, I_m)$ and $X_{j}^- \sim \mathcal{W}_{m}(v_m, \mathrm{AR}_{m}(0.4))$;
	\item  For $j\in \{5, 6\}$, $X_{j}^+ \sim \mathcal{W}_{m}(v_m, I_m)$ and $X_{j}^- \sim \mathcal{W}_{m}(v_m, \mathrm{HC}_{m}(0.5,\Lambda_m))$;	
	\item  For $j\in \{7, 8\}$, $X_{j}^+ \sim \mathcal{W}_{m}(v_m, \mathrm{AR}_{m}(-0.25))$ and $X_{j}^- \sim \mathcal{W}_{m}(v_m, \mathrm{AR}_{m}(0.25))$;

	\item  For $j\in \{9,10\}$, $X_{j}^+ \sim \mathcal{W}_{m}(v_m, \mathrm{HC}_{m}(0.2,\Lambda_m))$ and $X_{j}^- \sim \mathcal{W}_{m}(v_m, \mathrm{HC}_{m}(0.5,\Lambda_m))$;	

	\item For $j\in \{11,\dots,p\}$, both $X_{j}^+$ and $X_{j}^- \sim \mathcal{W}_{m}(v_m, \Sigma_m)$, where $\Sigma_m$ is randomly chosen from $I_m$, $\mathrm{AR}_{m}(-0.2)$, $\mathrm{AR}_{m}(0.5)$, or $\mathrm{HC}_{m}(0.5,\Lambda_m)$.
\end{enumerate}
In the above settings, $I_m$ denotes a $m\times m$ identity covariance matrix, and $\mathrm{AR}_{m}(\rho)$ denotes a $m \times m$  auto-regressive covariance matrix, with the $(k,l)$th element $\Sigma_{k,l}=\rho^{|k-l|}$. 
Similarly, $\mathrm{HC}_{m}(\rho, \Lambda_m)$ represents a $m \times m$ heterogeneous compound covariance matrix with $(k,l)$th element $\Sigma_{k,l}=\rho \lambda_k \lambda_l$ for $k \neq l$ and $\Sigma_{k,l}=\lambda_k^2$ for $k=l$, where $k,\ l =1,\dots,m$ and $\lambda_k$ is the $k$th element of the diagonal matrix $\Lambda_m$. 

We use three different metrics to measure the distance between $X_j^+$ and $X_j^-$, namely, 
the Euclidean metric \[d_{E}(X_j^+,X_j^-)=\|X_j^+ - X_j^-\|_F,\]
the Cholesky metric \[d_{C}(X_j^+,X_j^-)=\|\mathrm{chol}(X_j^+) - \mathrm{chol}(X_j^-) \|_F,\]
and the Log-Cholesky metric proposed by \citet{lin2019riemannian},  \[d_{LC}(X_j^+,X_j^-)=\left[\|\lfloor\mathrm{chol}(X_j^+)\rfloor - \lfloor\mathrm{chol}(X_j^-)\rfloor \|_F^2 + \| \log \mathbb{D}\{\mathrm{chol}(X_j^+)\} -  \log \mathbb{D}\{\mathrm{chol}(X_j^-)\}\|_F^2 \right]^{1/2}.\]
Here, $\|A\|_F=\{\mathrm{tr}(A^\T A)\}^{1/2}$ denotes the Frobenius norm of matrix $A$,
$\mathrm{chol}(A)$ represents the lower triangular matrix $L$ in the Cholesky decomposition $A=LL^\T$,
 $\lfloor A \rfloor$ denotes the strictly lower triangular part of $A$, and $\mathbb{D} (A)$ is the diagonal part of $A$. 

In our experiments, we consider $m=3$ and $m=5$ seperately,  with $v_{3}=v_{5}=10$, $\Lambda_3=\mathrm{diag}(1, 1.1, 1.2)$, and $\Lambda_5=\mathrm{diag}(1, 1.05, 1.1, 1.15, 1.2)$. For each setting, we fix $p=2000$ and generate $n=100$ observations. 
We control the FDR using the threshold given by Algorithm \ref{alg}, with significant level $\alpha=0.1$ and $0.2$, respectively.
We evaluate the performance of MK-Filter by computing the quantiles of MMS and the proportions $\mathcal{P}_{j}^{s}$ for $j=1, 3, 5, 7, 9$  based on $400$ replicates. 
The size of informative set is controlled by the threshold computed using Algorithm \ref{alg}. 
Additionally, we calculate the false discovery rate for each setting. Results are presented in Table~\ref{mkf:table2-1} and \ref{mkf:table2-2}. 
\input{doc/table-2.tex}

In both tables, we observe that the actual false discovery rates are below the corresponding nominal levels, indicating the effectiveness of  Algorithm \ref{alg}. 
The MMS of MK-Filter also closely approximate the true informative set size for all three metrics. 
Additionally, the proportions of informative features are selected significantly increase when using the Cholesky metric or Log-Cholesky metric, excerpt for $\mathcal{P}_1^s$. This can be understood by noticing that in Setting 1, we are comparing two $m \times m$ identity covariance matrices of different scales, which can also be seen as degenerate $m$-dimensional vectors. The Euclidean metric is more suitable  in this case because it does not involve misleading correltions among different elements. 
For all other settings, both Cholesky and Log-Cholesky metrics are preferable due to their attractive geometric properties.

%% file: doc/table-1.tex
\begin{table}[!h]
	\caption{Simulation results for distributional predictors.} 
		\begin{tabular}{lrrrrrrrrrrr}
			\hline
			\multicolumn{1}{c}{} & \multicolumn{1}{c}{\multirow{2}{*}{$\mathcal{P}_1^s$}} & \multicolumn{1}{c}{\multirow{2}{*}{$\mathcal{P}_2^s$}} & \multicolumn{1}{c}{\multirow{2}{*}{$\mathcal{P}_3^s$}} & \multicolumn{1}{c}{\multirow{2}{*}{$\mathcal{P}_4^s$}} & \multicolumn{1}{c}{\multirow{2}{*}{$\mathcal{P}_5^s$}} & \multicolumn{1}{c}{\multirow{2}{*}{$\mathcal{P}_6^s$}} & \multicolumn{1}{c}{\multirow{2}{*}{$\mathcal{P}_7^s$}} & \multicolumn{1}{c}{\multirow{2}{*}{$\mathcal{P}_8^s$}} & \multicolumn{3}{c}{Quantile of MMS} \\
			& \multicolumn{1}{c}{} & \multicolumn{1}{c}{} & \multicolumn{1}{c}{} & \multicolumn{1}{c}{} & \multicolumn{1}{c}{} & \multicolumn{1}{c}{} & \multicolumn{1}{c}{} & \multicolumn{1}{c}{} & $25\%$ & $50\%$ & $75\%$ \\ \hline
			Kolmogorov-Filter & 1 & 1 & 0.28 & 0.98 & 0 & 0.02 & 0.99 & 0.02 & 608 & 1311 & 2110 \\
			MV-SIS & 1 & 1 & 0.38 & 0.83 & 0 & 0.02 & 1 & 0.04 & 659 & 1154 & 1860 \\
			DC-SIS & 1 & 1 & 0.86 & 0.98 & 0.04 & 0.69 & 1 & 0.64 & 47 & 101 & 316 \\
			Ball-SIS & 1 & 1 & 1 & 1 & 0.24 & 0.94 & 1 & 0.09 & 77 & 212 & 534 \\
			MK-Filter & 1 & 1 & 0.96 & 0.96 & 0.93 & 1 & 1 & 0.96 & 8 & 8 & 10\\\hline
		\end{tabular}
	\label{mkf:table1}
\end{table}



%% file: doc/table-2.tex

\begin{table}[!h]
	\caption{Simulation results of the MF-Filter with different metrics for $\mathcal{S}_3^+$ features.} 
		\begin{tabular}{llccccccccc}
			\hline
			\multicolumn{1}{c}{} &  & \multicolumn{1}{l}{\multirow{2}{*}{FDR}} & \multirow{2}{*}{$\mathcal{P}_1^s$} & \multirow{2}{*}{$\mathcal{P}_3^s$} & \multirow{2}{*}{$\mathcal{P}_5^s$} & \multirow{2}{*}{$\mathcal{P}_7^s$} & \multirow{2}{*}{$\mathcal{P}_9^s$} & \multicolumn{3}{c}{Quantile of MMS} \\
			&  & \multicolumn{1}{l}{} &  &  &  &  &  & \multicolumn{1}{l}{$25\%$} & \multicolumn{1}{l}{$50\%$} & \multicolumn{1}{l}{$75\%$} \\ \cline{3-11} 
			& Euclidean & 0.053 & 0.871 & 0.7 & 0.279 & 0.686 & 0.307 & 11 & 14 & 23 \\
			$\alpha = 0.1$ & Cholesky & 0.03 & 0.493 & 0.8 & 0.3 & 0.714 & 0.457 & 10 & 10 & 11 \\
			& Log-Cholesky & 0.027 & 0.414 & 0.743 & 0.293 & 0.693 & 0.457 & 10 & 10 & 11 \\
			&  &  &  &  &  &  &  &  &  &  \\
			& Euclidean & 0.099 & 0.929 & 0.764 & 0.364 & 0.714 & 0.379 & 12 & 16 & 25 \\
			$\alpha = 0.2$ & Cholesky & 0.092 & 0.671 & 0.9 & 0.507 & 0.893 & 0.657 & 10 & 10 & 12 \\
			& Log-Cholesky & 0.088 & 0.643 & 0.893 & 0.507 & 0.914 & 0.629 & 10 & 10 & 12 \\ \hline
		\end{tabular}
	\label{mkf:table2-1}
\end{table}

\begin{table}[!h]
	\caption{Simulation results of the MF-Filter with different metrics for $\mathcal{S}_5^+$ features.} 
		\begin{tabular}{llccccccccc}
			\hline
			\multicolumn{1}{c}{} &  & \multicolumn{1}{l}{\multirow{2}{*}{FDR}} & \multirow{2}{*}{$\mathcal{P}_1^s$} & \multirow{2}{*}{$\mathcal{P}_3^s$} & \multirow{2}{*}{$\mathcal{P}_5^s$} & \multirow{2}{*}{$\mathcal{P}_7^s$} & \multirow{2}{*}{$\mathcal{P}_9^s$} & \multicolumn{3}{c}{Quantile of MMS} \\
			&  & \multicolumn{1}{l}{} &  &  &  &  &  & \multicolumn{1}{l}{$25\%$} & \multicolumn{1}{l}{$50\%$} & \multicolumn{1}{l}{$75\%$} \\ \cline{3-11} 
			& Euclidean & 0.072 & 0.929 & 0.857 & 0.557 & 0.786 & 0.557 & 10 & 10 & 10 \\
			$\alpha = 0.1$ & Cholesky & 0.038 & 0.7 & 0.857 & 0.4 & 0.821 & 0.614 & 10 & 10 & 10 \\
			& Log-Cholesky & 0.037 & 0.75 & 0.936 & 0.436 & 0.85 & 0.636 & 10 & 10 & 10 \\
			&  &  &  &  &  &  &  &  &  &  \\
			& Euclidean & 0.166 & 1 & 0.957 & 0.714 & 0.936 & 0.771 & 10 & 10 & 11 \\
			$\alpha = 0.2$ & Cholesky & 0.186 & 0.879 & 0.964 & 0.636 & 0.936 & 0.843 & 10 & 10 & 10 \\
			& Log-Cholesky & 0.169 & 0.9 & 0.986 & 0.764 & 0.95 & 0.864 & 10 & 10 & 10 \\ \hline
		\end{tabular}
	\label{mkf:table2-2}
\end{table}

%% file: doc/real-data.tex
\subsection{Study background}
Autism is a neurodevelopmental disorder characterized by deficits in social interaction and repetitive behaviors \citep{muhle2004genetics}. 
While its neurobiology largely remains unknown,  subtle alterations in brain regions, along with resulting abnormal functional connectivity patterns, are believed to play a crucial role in understanding autism \citep{ha2015characteristics, postema2019altered, dekhil2021identifying}. 
Researchers often explore these patterns by constructing covariance matrices of blood-oxygen-level dependent signals derived from brain activities in specific regions, using resting-state functional magnetic resonance imaging (fMRI) observations \citep{friston2011functional}. 
Such covariance matrices naturally form features in $\mathcal{S}^+$ when equipped with a suitable metric.

Traditional autism studies typically begin by selecting a set of $m$ regions of interest and then analyze their functional connectivities. 
However, our study suggests to take a different angle. 
We directly examine all possible $m \times m$ covariance matrices, representing interactions among arbitrary $m$ regions. 
This comprehensive exploration results in a large pool of potential features in metric spaces. 
We then apply the MK-Filter to identify informative features, specifically covariance matrices representing abnormal connectivities in autistic brains. 
The altered regions are identified by locating elements in these informative matrices. 
This inverse study procedure thoroughly investigates $m \times m$ functional connectivities without any omissions, aligning with the whole-brain analysis advocated by \citet{muller2011underconnected} and others. 
We ensure the significance of our findings by controlling the false discovery rate. This meticulous filtering process aims to uncover previously unknown abnormal brain connectivities, offering valuable insights for further investigation.



We analyze resting-state fMRI data from the Autism Brain Imaging Data Exchange (ABIDE, \citet{di2014autism}. 
Our main goal is to use the proposed MK-Filter to  identify distinct brain functional connectivities between individuals with autism and controls. 
These detected abnormal connectivities can further serve as informative features for detecting autism. 
Below, we describe our procedure in detail.

\subsection{Data processing}
	In this study, we utilize $149$ fMRI samples from  the ABIDE data collected at the NYU station, comprising $51$ samples from subjects with autism and $98$ from controls.
	All samples have been preprocessed using the Python package nilearn \citep{abraham2014machine}, employing the Conﬁgurable Pipeline for the Analysis of Connectomes (C-PAC, \citet{cpac2013}). 
	Each preprocessed sample is a tensor of dimension $61 \times 73 \times 61 \times 176$, representing a time series of length $176$, with each time index corresponding to an fMRI measurement voxel of size $61 \times 73 \times 61$.
	
	Based on the Harvard-Oxford cortical and subcortical structural atlases \citep{desikan2006automated}, the entire brain is segmented into $48$ cortical regions, denoted by $\mathcal{R}=\{R_1,\dots,R_{48}\}$. 
	Notice here we chose the Harvard-Oxford atlases for ease of illustration; however, other brain parcellation methods can also be utilized. 
	Given this parcellation, the preprocessed voxel-specific time series are divided into $48$ groups according to related regions and averaged, resulting in a $48 \times 176$ array of fMRI measurements. 


\subsection{Construct a feature pool of SPD matrices}
	For each subject $i \in \{1,\dots,149\}$, we compute a $48 \times 48$ covariance matrix $\Sigma_i$ using the  corresponding $48 \times 176$ fMRI data. 
	Such a matrix characterizes the functional connectivity of the entire brain and potentially contains information about abnormal brain connectivity related to autism, although might be obscured. 
	To detect subtle differences in local connectivities, we explore pairwise and triplewise connectivities between different brain regions, represented by covariance matrices of size $2 \times 2$ and $3 \times 3$, respectively. These matrices are easily obtained by selecting corresponding elements in $\Sigma_i$. 
	For instance, utilizing the interactions of the $j$th and $k$th rows and columns of $\Sigma_{i}$, we derive a $2\times 2$ sub-matrix denoted as $\Sigma_{i}(R_j,R_k)$, characterizing the functional connectivity between regions $R_j$ and $R_k$, where $R_j, R_k \in \mathcal{R}$. 
	
\begin{figure}[!h]
	\centering
	\subcaptionbox{The whole-brain covariance matrix $\Sigma$. }{\includegraphics[width=0.54\textwidth]{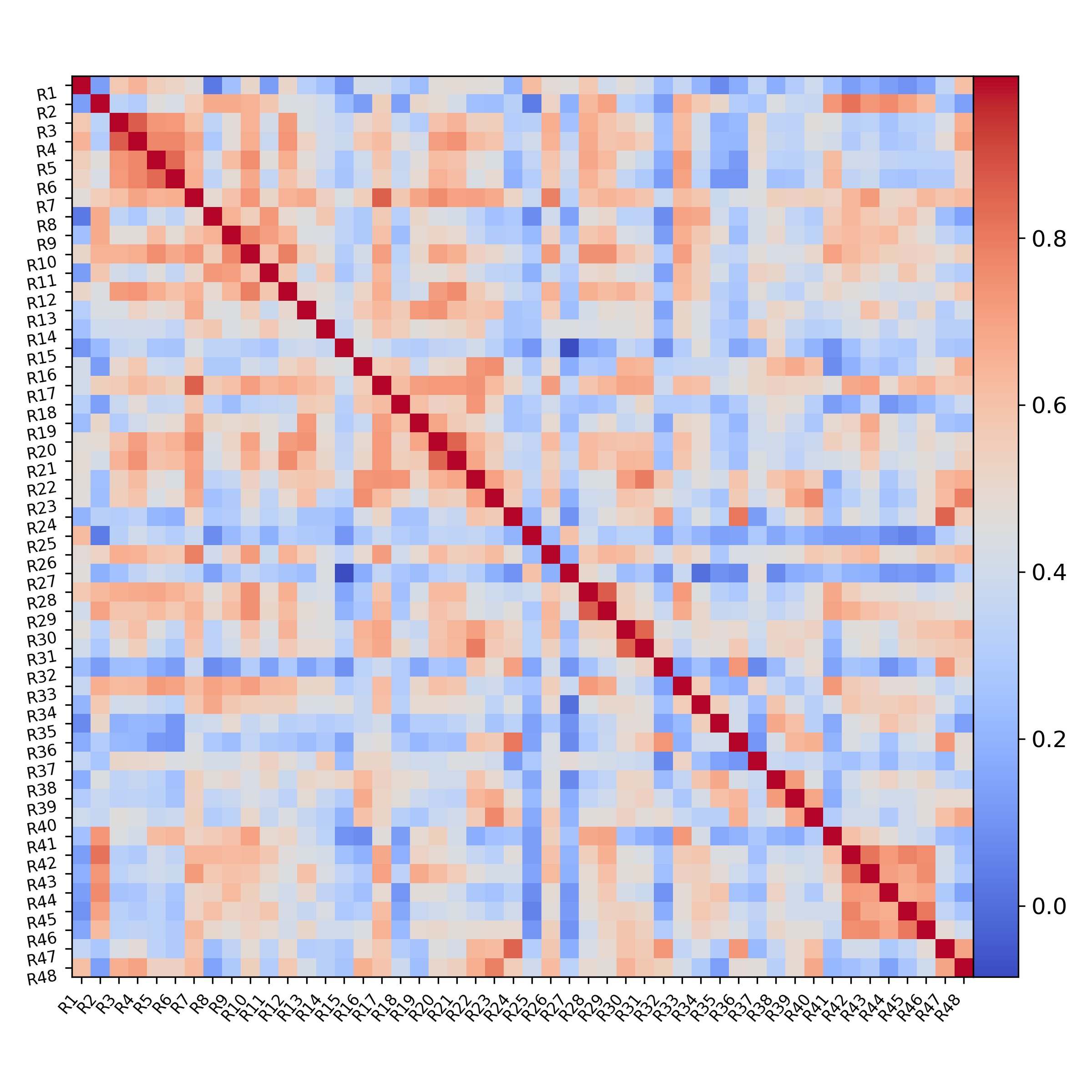}}
	\subcaptionbox{The sub-matrices of $\Sigma$ in $\mathcal{S}_2^+$ and $\mathcal{S}_3^+$. }{\includegraphics[width=0.45\textwidth]{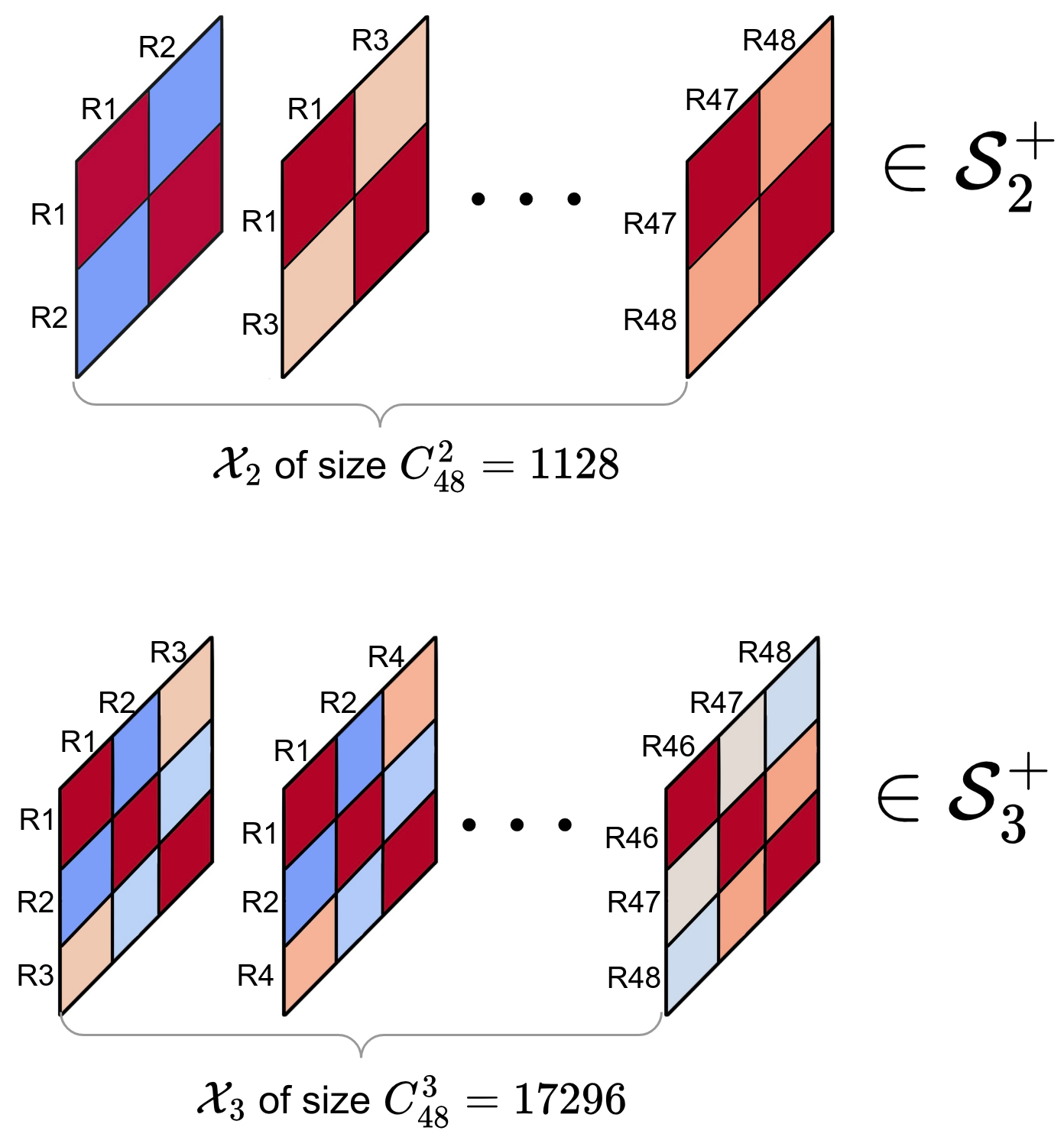}}
	\caption{A feature pool derived from the whole-brain covariance for autism detection. Panel (a) displays the covariance matrix of $48$ brain regions of interest, utilizing fMRI data from a chosen subject. Panel (b) illustrates $C_{48}^2$ random objects in $\mathcal{S}_2^+$ and $C_{48}^3$ objects $\mathcal{S}_3^+$ as potential features, which represent combinations of sub-matrices derived from $\Sigma$, with size $2 \times 2$ and $3 \times 3$, respectively.}
	\label{fig:data-process}	
\end{figure}

	Treating $\Sigma_{i}(R_j,R_k)$ and $\Sigma_{i^\prime}(R_j,R_k)$, calculated from the subject $i$ with autism and $i^\prime$ the control, as random objects in $\mathcal{S}_2^+$,  various distances can be employed to compare their similarities. 
	Here, we employ the log-Cholesky metric $d_{LC}(\cdot,\cdot)$ for its nice properties \citep{lin2019riemannian}. 
	Roughly speaking, when $d_{LC}\{\Sigma_{i}(R_j,R_k), \Sigma_{i^\prime}(R_j,R_k)\}$ is notably large, it suggests significant differences between $\Sigma_{i}(R_j,R_k)$ and $\Sigma_{i^\prime}(R_j,R_k)$, implying that $\Sigma(R_j,R_k)$ might be an informative feature for detecting autism.
	This inspires us to examine all such $2\times 2$ matrices. 
	Considering combinations of selecting $2$ regions from the set $\mathcal{R}$ of $48$ elements, we obtain $C_{48}^2=1128$ covariance matrices, resulting in a collection of features $\mathcal{X}_2=\{\Sigma(R_j,R_k) \in \mathcal{S}_2^+:  R_j, R_k \in \mathcal{R}, \text{and}\  j \neq k\}$.
	Similarly, we take $\Sigma(R_j, R_k, R_l)$ to describe the triplewise connectivies among regions $R_{j}$, $R_{k}$, and $R_{l}$. 
	Combining all such $3 \times 3$ sub-matrices also constructs a feature pool $\mathcal{X}_3=\{\Sigma(R_j, R_k, R_l) \in \mathcal{S}_3^+:  R_j, R_k, R_l \in \mathcal{R}, \text{and}\  j \neq k \neq l\}$ of size $C_{48}^3=17296$.  
	The whole procedure is illustrated in Figure~\ref{fig:data-process}.
	Detecting autism based on these large-scale features is challenging, given the limited sample size of $n=149$.  
	This motivates us to first obtain an informative feature set.

\subsection{Abnormal connectivities detection}\label{step-detect}
	We apply the MK-Filter to $(\mathcal{X}_2,d_{LC})$ and $(\mathcal{X}_3,d_{LC})$, respectively. 
	With significant level $\alpha=0.1$, a few covariance matrices are identified as informative features that effectively differentiate functional connectivities between individuals with and without autism, as shown in Table~\ref{table:real-features}.
	\begin{table}[!h]
		\centering
		\caption{Informative features obtained by MK-Filter.} 
			\begin{tabular}{lcl}
				\hline
				\multicolumn{1}{c}{} & Informative features & \multicolumn{1}{c}{Related regions} \\ \cline{2-3} 
				&  & \multirow{8}{*}{\begin{tabular}[c]{@{}l@{}}$R_1$: Frontal Pole;\\ $R_{8}$: Temporal Pole;\\ $R_{11}$: Middle Temporal Gyrus (anterior division);\\ $R_{12}$: MiddleTemporal Gyrus (posterior division);\\ $R_{13}$: Middle Temporal Gyrus (temporooccipital part);\\ $R_{32}$: Cuneal Cortex;\\ $R_{34}$: Parahippocampal Gyrus (anterior division);\\ $R_{35}$: Parahippocampal Gyrus (posterior division).\end{tabular}} \\
				$\mathcal{S}_{2}^+$ & $\Sigma (R_1,R_{11})$ &  \\
				& $\Sigma (R_{8},R_{32})$ &  \\
				&  &  \\
				$\mathcal{S}_{3}^+$ & $\Sigma (R_1,R_{11},R_{12})$ &  \\
				& $\Sigma (R_1,R_{12},R_{35})$ &  \\
				& $\Sigma (R_1,R_{11},R_{34})$ &  \\
				& $\Sigma (R_1,R_{11},R_{13})$ &  \\ \hline
			\end{tabular}
		\label{table:real-features}
	\end{table}
	
	The last column of Table~\ref{table:real-features} lists significantly altered brain regions in autism compared with controls. 
	Some, such as frontal pole, middle temporal gyrus, and parahippocampal gyrus, have been widely reported in MRI-based autism studies to show reduced connectivity to other regions (see \citet{rane2015connectivity} for a detailed review). 
	Others, like temporal pole and cuneal cortex, have been less frequently reported and may warrant future investigation. 
	Importantly, our findings extend beyond individual brain region analyses, directly identifying abnormal interactions among brain regions in autism. 
	In this case study, the abnormal connectivities are filtered from an exhaustive analysis of $1128$ pairwise and $17296$ triplewise interactions, covering all conceivable combinations. 
	Consequently, the reported set of informative features is expected to capture all potentially significant pairwise and triplewise connectivities for detecting autism with a high probability, while controling the false discovery rate at $\alpha=0.1$. 
	By precisely identifying the involved brain regions and their related connectivities, our results provide a helpful understanding of the disorder's neural mechanisms.		
		
\begin{figure}[!ht]
		\centering
		\includegraphics[width=0.8\textwidth]{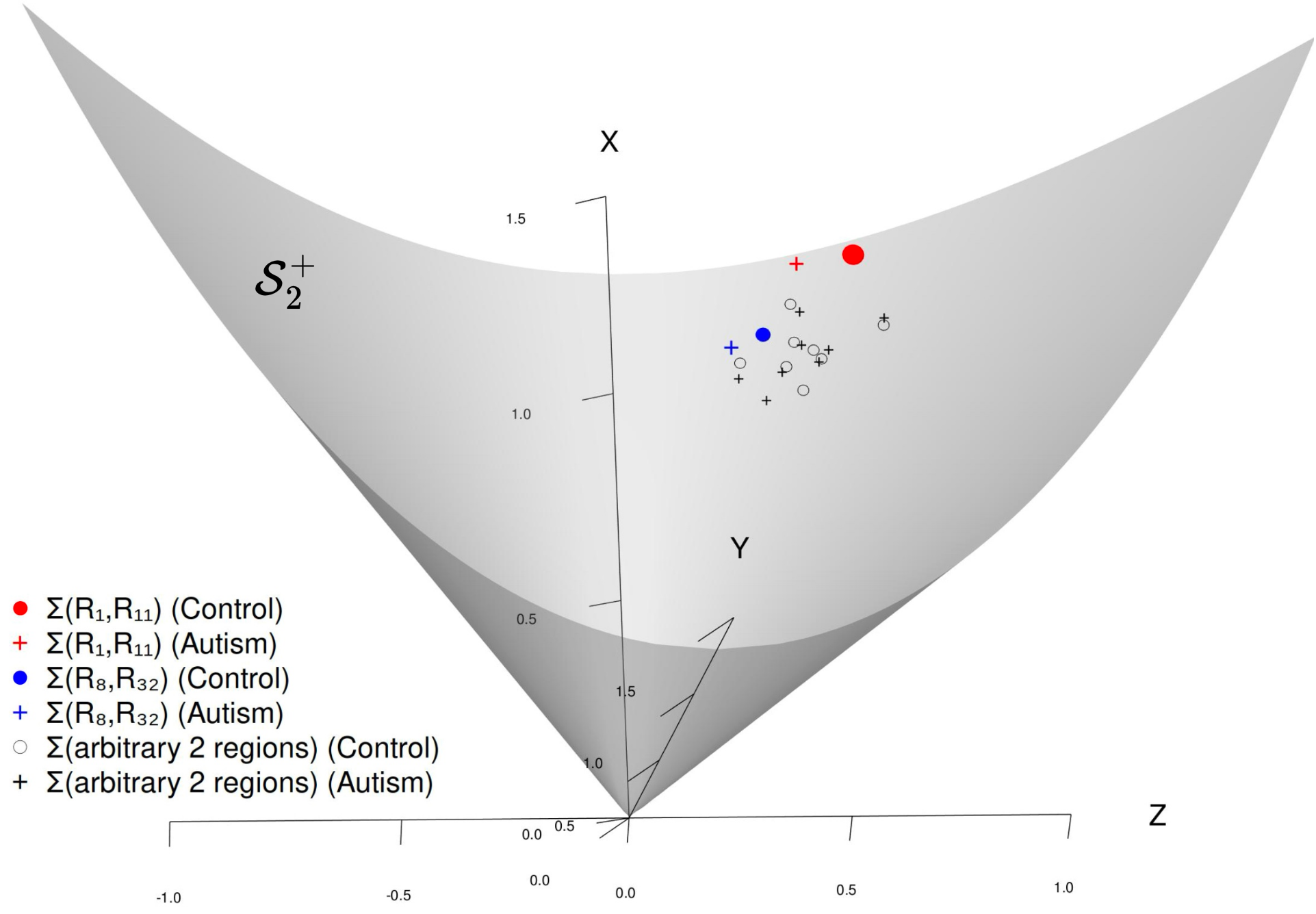}
		\caption{Informative features (marked with red or blue colors) within $\mathcal{S}_2^+$ exhibit more distinct values between individuals with austism (denoted by ``+'') and controls (denoted by circles), compared to uninformative counterparts.}
		\label{fig:S2}
	\end{figure}

	The findings presented in Table~\ref{table:real-features} can also be explained geometrically.
	Notice the matrix $\Sigma (R_j,R_k)$ takes the form: 
	\begin{equation}\label{fm:S2-condition}  
		\Sigma (R_j,R_k)=\left(\begin{array}{ll}a &\quad b \\ b & \quad c\end{array}\right), \quad a c-b^2>0, \quad a>0,
	\end{equation}
	representing a point $(a,b,c)^\T \in \mathbb{R}^3$.
	All such matrices form a cone $\mathcal{S}_2^+$ in $\mathbb{R}^3$, and $\mathcal{X}_2$ is a set of points  situated in this cone. 
	For each feature $\Sigma(R_j,R_k) \in \mathcal{X}_2$, we compute its sample averages based on different sample classes (i.e., autism or control),  resulting in a pair of points within this cone. 
	In Figure~\ref{fig:S2}, red and blue points represent informative features $\Sigma (R_1,R_{11})$ and $\Sigma (R_8,R_{32})$, respectively, while the black points correspond to uninformative features randomly chosen from $\mathcal{X}_2$. 
	Here, the symbol ``+'' denotes the autism class, and the dot represents the control. 
 	By comparing the pairs of black points to the red (or blue) points, we observe a larger distance within the latter, indicating more distinct characteristics between autism and control.
	The use of $\mathcal{S}_2^+$ serves as an intuitive visual example, but similar conclusions apply to features in $\mathcal{S}_3^+$. 
	
\subsection{Classification}
	In the final step, we evaluate the prediction performance using the informative features identified by MK-Filter. 
	The $k$-Nearest Neighbors ($k$-NN) classifier is a natural choice for binary classification within a metric space. 
	It predicts the label of a new sample $x_0$ based on the majority vote of its $k$ closest neighbors, determined by measusing distances or metrics between $x_0$ and exisiting samples. 
	In our experiments, we employ the $k$-NN classifier on two different feature sets: the whole-brain covariance matrix of size $48 \times 48$, or the informative features identified by the MK-Filter in Step~\ref{step-detect}, utilizing the Log-Cholesky metric. 
	
	In the second scenario, determining the nearest points of  $x_0$  requires considering multiple features, posing a multivariate ranking challenge. 
	We propose two strategies to address this:  the first one is ``voting'', which applies the $k$-NN classifier on each feature separately and then combines the results via voting to obtain the prediction; 
	the second approach, termed ``merging'', works by assigning a new metric on the product space of the informative features, thus allowing for a univariate ranking. 
	Specifically, for the tuple $X=(X_1,\dots,X_s)$ comprising informative features $X_j \in (\mathcal{M}_j, d_j)$, $j =1,\dots,s$, we define a new metric space $(\mathcal{M},d)$. 
	Here, $\mathcal{M}=\prod_{j=1}^{s} \mathcal{M}_j$ represents the Cartesian product of $\mathcal{M}_j$, and $d(u,v)=\{\sum_{j=1}^{s} d_{1}^{2}(u_j, v_s)\}^{1/2}$ for $u, v \in \mathcal{M}$ is the corresponding metric.
	The $k$-NN classifier can then be applied using this metric $d(\cdot,\cdot)$.
	
	We conduct experiments for each setting with $k$ values $3$, $6$, and $9$. 
	In each experiment, the samples are randomly split into training and testing sets in a ratio of $7:3$, and the classification performance on the testing set is recorded. 
	Table~\ref{table:real-classification} summarizes the averaged results from $400$ replicates, with standard errors shown in parentheses.  
	It's evident that the $k$-NN classifier using identified informative features exhibits significant improvements for both strategies. 

\begin{table}[!h]
	\centering
	\caption{The performance of $k$-NN classifier using different features.} 
		\begin{tabular}{crlll}
			\hline
			$k$ & \multicolumn{1}{c}{Feature} & \multicolumn{1}{c}{Accuracy} & \multicolumn{1}{c}{Recall rate} & \multicolumn{1}{c}{F1-Score} \\ \hline
			& \multicolumn{1}{l}{Whole-brain} & 0.652 (0.06) & 0.172 (0.09) & 0.241 (0.10) \\
			3 & MK-Filter + merging & 0.729 (0.06) & 0.523 (0.11) & 0.559 (0.09) \\
			& + voting & 0.720 (0.05) & 0.559 (0.11) & 0.567 (0.08) \\
			&  & \multicolumn{1}{r}{} & \multicolumn{1}{r}{} & \multicolumn{1}{r}{} \\
			& \multicolumn{1}{l}{Whole-brain} & 0.666 (0.06) & 0.270 (0.12) & 0.343 (0.12) \\
			6 & MK-Filter + merging & 0.730 (0.05) & 0.515 (0.11) & 0.559 (0.09) \\
			& + voting & 0.708 (0.06) & 0.636 (0.10) & 0.594 (0.08) \\
			&  & \multicolumn{1}{r}{} & \multicolumn{1}{r}{} & \multicolumn{1}{r}{} \\
			& \multicolumn{1}{l}{Whole-brain} & 0.680 (0.06) & 0.156 (0.09) & 0.239 (0.11) \\
			9 & MK-Filter + merging & 0.723 (0.06) & 0.497 (0.11) & 0.546 (0.10) \\
			& + voting & 0.723 (0.06) & 0.505 (0.13) & 0.549 (0.09) \\ \hline
		\end{tabular}
	\label{table:real-classification}
\end{table}

%% file: MK-Filter-Arxiv.bbl
\begin{thebibliography}{45}
\providecommand{\natexlab}[1]{#1}
\providecommand{\url}[1]{\texttt{#1}}
\expandafter\ifx\csname urlstyle\endcsname\relax
  \providecommand{\doi}[1]{doi: #1}\else
  \providecommand{\doi}{doi: \begingroup \urlstyle{rm}\Url}\fi

\bibitem[Abraham et~al.(2014)Abraham, Pedregosa, Eickenberg, Gervais, Mueller,
  Kossaifi, Gramfort, Thirion, and Varoquaux]{abraham2014machine}
Alexandre Abraham, Fabian Pedregosa, Michael Eickenberg, Philippe Gervais,
  Andreas Mueller, Jean Kossaifi, Alexandre Gramfort, Bertrand Thirion, and
  Ga{\"e}l Varoquaux.
\newblock Machine learning for neuroimaging with scikit-learn.
\newblock \emph{Frontiers in neuroinformatics}, 8:\penalty0 14, 2014.

\bibitem[Arsigny et~al.(2007)Arsigny, Fillard, Pennec, and
  Ayache]{arsigny2007geometric}
Vincent Arsigny, Pierre Fillard, Xavier Pennec, and Nicholas Ayache.
\newblock Geometric means in a novel vector space structure on symmetric
  positive-definite matrices.
\newblock \emph{SIAM journal on matrix analysis and applications}, 29\penalty0
  (1):\penalty0 328--347, 2007.

\bibitem[Benjamini and Hochberg(1995)]{benjamini1995controlling}
Yoav Benjamini and Yosef Hochberg.
\newblock Controlling the false discovery rate: a practical and powerful
  approach to multiple testing.
\newblock \emph{Journal of the Royal statistical society: series B
  (Methodological)}, 57\penalty0 (1):\penalty0 289--300, 1995.

\bibitem[Bhattacharjee and M{\"u}ller(2023)]{bhattacharjee2023single}
Satarupa Bhattacharjee and Hans-Georg M{\"u}ller.
\newblock Single index fr{\'e}chet regression.
\newblock \emph{The Annals of Statistics}, 51\penalty0 (4):\penalty0
  1770--1798, 2023.

\bibitem[Billera et~al.(2001)Billera, Holmes, and
  Vogtmann]{billera2001geometry}
Louis~J Billera, Susan~P Holmes, and Karen Vogtmann.
\newblock Geometry of the space of phylogenetic trees.
\newblock \emph{Advances in Applied Mathematics}, 27\penalty0 (4):\penalty0
  733--767, 2001.

\bibitem[Chaudhuri and Dasgupta(2014)]{chaudhuri2014rates}
Kamalika Chaudhuri and Sanjoy Dasgupta.
\newblock Rates of convergence for nearest neighbor classification.
\newblock \emph{Advances in Neural Information Processing Systems}, 27, 2014.

\bibitem[Chen et~al.(2023)Chen, Lin, and M{\"u}ller]{chen2023wasserstein}
Yaqing Chen, Zhenhua Lin, and Hans-Georg M{\"u}ller.
\newblock Wasserstein regression.
\newblock \emph{Journal of the American Statistical Association}, 118\penalty0
  (542):\penalty0 869--882, 2023.

\bibitem[Cover and Hart(1967)]{cover1967nearest}
Thomas Cover and Peter Hart.
\newblock Nearest neighbor pattern classification.
\newblock \emph{IEEE transactions on information theory}, 13\penalty0
  (1):\penalty0 21--27, 1967.

\bibitem[Craddock et~al.(2013)Craddock, Sikka, Cheung, Khanuja, Ghosh, Yan, Li,
  Lurie, Vogelstein, Burns, Colcombe, Mennes, Kelly, Di~Martino, Castellanos,
  and Milham]{cpac2013}
Cameron Craddock, Sharad Sikka, Brian Cheung, Ranjeet Khanuja, Satrajit~S
  Ghosh, Chaogan Yan, Qingyang Li, Daniel Lurie, Joshua Vogelstein, Randal
  Burns, Stanley Colcombe, Maarten Mennes, Clare Kelly, Adriana Di~Martino,
  Francisco~Xavier Castellanos, and Michael Milham.
\newblock Towards automated analysis of connectomes: The {Configurable Pipeline
  for the Analysis of Connectomes (C-PAC)}.
\newblock \emph{Frontiers in Neuroinformatics}, \penalty0 (42), 2013.
\newblock ISSN 1662-5196.

\bibitem[Cui et~al.(2015)Cui, Li, and Zhong]{cui2015model}
Hengjian Cui, Runze Li, and Wei Zhong.
\newblock Model-free feature screening for ultrahigh dimensional discriminant
  analysis.
\newblock \emph{Journal of the American Statistical Association}, 110\penalty0
  (510):\penalty0 630--641, 2015.

\bibitem[Dekhil et~al.(2021)Dekhil, Shalaby, Soliman, Mahmoud, Kong, Barnes,
  Elmaghraby, and El-Baz]{dekhil2021identifying}
Omar Dekhil, Ahmed Shalaby, Ahmed Soliman, Ali Mahmoud, Maiying Kong, Gregory
  Barnes, Adel Elmaghraby, and Ayman El-Baz.
\newblock Identifying brain areas correlated with ados raw scores by studying
  altered dynamic functional connectivity patterns.
\newblock \emph{Medical Image Analysis}, 68:\penalty0 101899, 2021.

\bibitem[Desikan et~al.(2006)Desikan, S{\'e}gonne, Fischl, Quinn, Dickerson,
  Blacker, Buckner, Dale, Maguire, Hyman, et~al.]{desikan2006automated}
Rahul~S Desikan, Florent S{\'e}gonne, Bruce Fischl, Brian~T Quinn, Bradford~C
  Dickerson, Deborah Blacker, Randy~L Buckner, Anders~M Dale, R~Paul Maguire,
  Bradley~T Hyman, et~al.
\newblock An automated labeling system for subdividing the human cerebral
  cortex on mri scans into gyral based regions of interest.
\newblock \emph{Neuroimage}, 31\penalty0 (3):\penalty0 968--980, 2006.

\bibitem[Di~Martino et~al.(2014)Di~Martino, Yan, Li, Denio, Castellanos,
  Alaerts, Anderson, Assaf, Bookheimer, Dapretto, et~al.]{di2014autism}
Adriana Di~Martino, Chao-Gan Yan, Qingyang Li, Erin Denio, Francisco~X
  Castellanos, Kaat Alaerts, Jeffrey~S Anderson, Michal Assaf, Susan~Y
  Bookheimer, Mirella Dapretto, et~al.
\newblock The autism brain imaging data exchange: towards a large-scale
  evaluation of the intrinsic brain architecture in autism.
\newblock \emph{Molecular psychiatry}, 19\penalty0 (6):\penalty0 659--667,
  2014.

\bibitem[Dryden et~al.(2009)Dryden, Koloydenko, and Zhou]{10.1214/09-AOAS249}
Ian~L. Dryden, Alexey Koloydenko, and Diwei Zhou.
\newblock {Non-Euclidean statistics for covariance matrices, with applications
  to diffusion tensor imaging}.
\newblock \emph{The Annals of Applied Statistics}, 3\penalty0 (3):\penalty0
  1102 -- 1123, 2009.
\newblock \doi{10.1214/09-AOAS249}.
\newblock URL \url{https://doi.org/10.1214/09-AOAS249}.

\bibitem[Dubey and M{\"u}ller(2020)]{dubey2020functional}
Paromita Dubey and Hans-Georg M{\"u}ller.
\newblock Functional models for time-varying random objects.
\newblock \emph{Journal of the Royal Statistical Society Series B: Statistical
  Methodology}, 82\penalty0 (2):\penalty0 275--327, 2020.

\bibitem[Dvoretzky et~al.(1956)Dvoretzky, Kiefer, and
  Wolfowitz]{dvoretzky1956asymptotic}
Aryeh Dvoretzky, Jack Kiefer, and Jacob Wolfowitz.
\newblock Asymptotic minimax character of the sample distribution function and
  of the classical multinomial estimator.
\newblock \emph{The Annals of Mathematical Statistics}, pages 642--669, 1956.

\bibitem[Fan and Lv(2008)]{fan2008sure}
Jianqing Fan and Jinchi Lv.
\newblock Sure independence screening for ultrahigh dimensional feature space.
\newblock \emph{Journal of the Royal Statistical Society Series B: Statistical
  Methodology}, 70\penalty0 (5):\penalty0 849--911, 2008.

\bibitem[Fan and Lv(2018)]{fan2018sure}
Jianqing Fan and Jinchi Lv.
\newblock Sure independence screening.
\newblock \emph{Wiley StatsRef: Statistics Reference Online}, 2018.

\bibitem[FAN and SONG(2010)]{fan2010sure}
JIANQING FAN and RUI SONG.
\newblock Sure independence screening in generalized linear models with
  np-dimensionality.
\newblock \emph{The Annals of Statistics}, 38\penalty0 (6):\penalty0
  3567--3604, 2010.

\bibitem[Fournier and Guillin(2015)]{fournier2015rate}
Nicolas Fournier and Arnaud Guillin.
\newblock On the rate of convergence in wasserstein distance of the empirical
  measure.
\newblock \emph{Probability Theory and Related Fields}, 162\penalty0
  (3):\penalty0 707--738, 2015.

\bibitem[Friston(2011)]{friston2011functional}
Karl~J Friston.
\newblock Functional and effective connectivity: a review.
\newblock \emph{Brain connectivity}, 1\penalty0 (1):\penalty0 13--36, 2011.

\bibitem[Gottlieb et~al.(2014)Gottlieb, Kontorovich, and
  Krauthgamer]{gottlieb2014efficient}
Lee-Ad Gottlieb, Aryeh Kontorovich, and Robert Krauthgamer.
\newblock Efficient classification for metric data.
\newblock \emph{IEEE Transactions on Information Theory}, 60\penalty0
  (9):\penalty0 5750--5759, 2014.

\bibitem[Guo et~al.(2023)Guo, Ren, Zou, and Li]{guo2023threshold}
Xu~Guo, Haojie Ren, Changliang Zou, and Runze Li.
\newblock Threshold selection in feature screening for error rate control.
\newblock \emph{Journal of the American Statistical Association}, 118\penalty0
  (543):\penalty0 1773--1785, 2023.

\bibitem[Ha et~al.(2015)Ha, Sohn, Kim, Sim, and Cheon]{ha2015characteristics}
Sungji Ha, In-Jung Sohn, Namwook Kim, Hyeon~Jeong Sim, and Keun-Ah Cheon.
\newblock Characteristics of brains in autism spectrum disorder: structure,
  function and connectivity across the lifespan.
\newblock \emph{Experimental neurobiology}, 24\penalty0 (4):\penalty0 273,
  2015.

\bibitem[Kleiber and Pervin(1969)]{kleiber1969generalized}
Martin Kleiber and William~J Pervin.
\newblock A generalized banach-mazur theorem.
\newblock \emph{Bulletin of The Australian Mathematical Society}, 1\penalty0
  (2):\penalty0 169--173, 1969.

\bibitem[Kontorovich et~al.(2018)Kontorovich, Sabato, and
  Urner]{kontorovich2018active}
Aryeh Kontorovich, Sivan Sabato, and Ruth Urner.
\newblock Active nearest-neighbor learning in metric spaces.
\newblock \emph{Journal of Machine Learning Research}, 18\penalty0
  (195):\penalty0 1--38, 2018.

\bibitem[Lei(2020)]{lei2020convergence}
Jing Lei.
\newblock Convergence and concentration of empirical measures under wasserstein
  distance in unbounded functional spaces.
\newblock \emph{Bernoulli}, 26\penalty0 (1):\penalty0 767--798, 2020.

\bibitem[Lim et~al.(2021)Lim, Wong, and Ye]{lim2021grassmannian}
Lek-Heng Lim, Ken Sze-Wai Wong, and Ke~Ye.
\newblock The grassmannian of affine subspaces.
\newblock \emph{Foundations of Computational Mathematics}, 21:\penalty0
  537--574, 2021.

\bibitem[Lin and Lin(2023)]{lin2023logistic}
Yinan Lin and Zhenhua Lin.
\newblock Logistic regression and classification with non-euclidean covariates.
\newblock \emph{arXiv preprint arXiv:2302.11746}, 2023.

\bibitem[Lin(2019)]{lin2019riemannian}
Zhenhua Lin.
\newblock Riemannian geometry of symmetric positive definite matrices via
  cholesky decomposition.
\newblock \emph{SIAM Journal on Matrix Analysis and Applications}, 40\penalty0
  (4):\penalty0 1353--1370, 2019.

\bibitem[Mai and Zou(2013)]{mai2013kolmogorov}
Qing Mai and Hui Zou.
\newblock The kolmogorov filter for variable screening in high-dimensional
  binary classification.
\newblock \emph{Biometrika}, 100\penalty0 (1):\penalty0 229--234, 2013.

\bibitem[Mardia et~al.(2000)Mardia, Jupp, and Mardia]{mardia2000directional}
Kanti~V Mardia, Peter~E Jupp, and KV~Mardia.
\newblock \emph{Directional statistics}, volume~2.
\newblock Wiley Online Library, 2000.

\bibitem[McCormack and Hoff(2021)]{mccormack2021equivariant}
Andrew McCormack and Peter Hoff.
\newblock Equivariant estimation of fr$\backslash$'echet means.
\newblock \emph{arXiv preprint arXiv:2104.03397}, 2021.

\bibitem[Muhle et~al.(2004)Muhle, Trentacoste, and Rapin]{muhle2004genetics}
Rebecca Muhle, Stephanie~V Trentacoste, and Isabelle Rapin.
\newblock The genetics of autism.
\newblock \emph{Pediatrics}, 113\penalty0 (5):\penalty0 e472--e486, 2004.

\bibitem[M{\"u}ller et~al.(2011)M{\"u}ller, Shih, Keehn, Deyoe, Leyden, and
  Shukla]{muller2011underconnected}
Ralph-Axel M{\"u}ller, Patricia Shih, Brandon Keehn, Janae~R Deyoe, Kelly~M
  Leyden, and Dinesh~K Shukla.
\newblock Underconnected, but how? a survey of functional connectivity mri
  studies in autism spectrum disorders.
\newblock \emph{Cerebral cortex}, 21\penalty0 (10):\penalty0 2233--2243, 2011.

\bibitem[Pan et~al.(2019)Pan, Wang, Xiao, and Zhu]{pan2018generic}
Wenliang Pan, Xueqin Wang, Weinan Xiao, and Hongtu Zhu.
\newblock A generic sure independence screening procedure.
\newblock \emph{Journal of the American Statistical Association}, 2019.

\bibitem[Petersen and M{\"u}ller(2019)]{petersen2019frechet}
Alexander Petersen and Hans-Georg M{\"u}ller.
\newblock Fr{\'e}chet regression for random objects with euclidean predictors.
\newblock 2019.

\bibitem[Postema et~al.(2019)Postema, Van~Rooij, Anagnostou, Arango, Auzias,
  Behrmann, Filho, Calderoni, Calvo, Daly, et~al.]{postema2019altered}
Merel~C Postema, Daan Van~Rooij, Evdokia Anagnostou, Celso Arango, Guillaume
  Auzias, Marlene Behrmann, Geraldo~Busatto Filho, Sara Calderoni, Rosa Calvo,
  Eileen Daly, et~al.
\newblock Altered structural brain asymmetry in autism spectrum disorder in a
  study of 54 datasets.
\newblock \emph{Nature communications}, 10\penalty0 (1):\penalty0 4958, 2019.

\bibitem[Rane et~al.(2015)Rane, Cochran, Hodge, Haselgrove, Kennedy, and
  Frazier]{rane2015connectivity}
Pallavi Rane, David Cochran, Steven~M Hodge, Christian Haselgrove, David~N
  Kennedy, and Jean~A Frazier.
\newblock Connectivity in autism: a review of mri connectivity studies.
\newblock \emph{Harvard review of psychiatry}, 23\penalty0 (4):\penalty0
  223--244, 2015.

\bibitem[von Luxburg and Bousquet(2004)]{von2004distance}
Ulrike von Luxburg and Olivier Bousquet.
\newblock Distance-based classification with lipschitz functions.
\newblock \emph{J. Mach. Learn. Res.}, 5\penalty0 (Jun):\penalty0 669--695,
  2004.

\bibitem[Wang et~al.(2021)Wang, Zhu, Pan, Zhu, and
  Zhang]{wang2021nonparametric}
Xueqin Wang, Jin Zhu, Wenliang Pan, Junhao Zhu, and Heping Zhang.
\newblock Nonparametric statistical inference via metric distribution function
  in metric spaces.
\newblock \emph{arXiv e-prints}, pages arXiv--2107, 2021.

\bibitem[Wang et~al.(2023)Wang, Zhu, Pan, Zhu, and
  Zhang]{wang2023nonparametric}
Xueqin Wang, Jin Zhu, Wenliang Pan, Junhao Zhu, and Heping Zhang.
\newblock Nonparametric statistical inference via metric distribution function
  in metric spaces.
\newblock \emph{Journal of the American Statistical Association}, pages 1--13,
  2023.

\bibitem[Ying and Yu(2022)]{ying2022frechet}
Chao Ying and Zhou Yu.
\newblock Fr{\'e}chet sufficient dimension reduction for random objects.
\newblock \emph{Biometrika}, 109\penalty0 (4):\penalty0 975--992, 2022.

\bibitem[Zhang et~al.(2023)Zhang, Xue, and Li]{zhang2023dimension}
Qi~Zhang, Lingzhou Xue, and Bing Li.
\newblock Dimension reduction for fr{\'e}chet regression.
\newblock \emph{Journal of the American Statistical Association}, \penalty0
  (just-accepted):\penalty0 1--27, 2023.

\bibitem[Zhu et~al.(2011)Zhu, Li, Li, and Zhu]{zhu2011model}
Li-Ping Zhu, Lexin Li, Runze Li, and Li-Xing Zhu.
\newblock Model-free feature screening for ultrahigh-dimensional data.
\newblock \emph{Journal of the American Statistical Association}, 106\penalty0
  (496):\penalty0 1464--1475, 2011.

\end{thebibliography}
